\documentclass[twocolumn]{aastex63}

\graphicspath{{./}{figures/}}
\usepackage{pifont}
\usepackage{multirow}
\usepackage{amsmath}
\usepackage{color}

\begin{document}

\title{Population Properties of Gravitational-Wave Neutron Star--Black Hole Mergers }

\author[0000-0002-9195-4904]{Jin-Ping Zhu}
\affil{Department of Astronomy, School of Physics, Peking University, Beijing 100871, China; \url{zhujp@pku.edu.cn}}

\author[0000-0002-9188-5435]{Shichao Wu}
\affiliation{Max-Planck-Institut f{\"u}r Gravitationsphysik (Albert-Einstein-Institut), D-30167 Hannover, Germany; \url{shichao.wu@aei.mpg.de}}
\affiliation{Leibniz Universit{\"a}t Hannover, D-30167 Hannover, Germany}

\author[0000-0002-2956-8367]{Ying Qin}
\affiliation{Department of Physics, Anhui Normal University, Wuhu, Anhui, 241000, China}

\author[0000-0002-9725-2524]{Bing Zhang}
\affiliation{Nevada Center for Astrophysics, University of Nevada, Las Vegas, NV 89154, USA; \url{bing.zhang@unlv.edu}}
\affiliation{Department of Physics and Astronomy, University of Nevada, Las Vegas, NV 89154, USA}

\author[0000-0002-3100-6558]{He Gao}
\affiliation{Department of Astronomy, Beijing Normal University, Beijing 100875, China; \url{zjcao@amt.ac.cn}}

\author[0000-0002-1932-7295]{Zhoujian Cao}
\affiliation{Department of Astronomy, Beijing Normal University, Beijing 100875, China; \url{zjcao@amt.ac.cn}}

\begin{abstract}

Over the course of the third observing run of LIGO-Virgo-KAGRA Collaboration, several gravitational-wave (GW) neutron star--black hole (NSBH) candidates have been announced. By assuming that these candidates are {real signals} of astrophysical origins, we analyze the population properties of the mass and spin distributions for GW NSBH mergers. We find that the primary BH mass distribution of NSBH systems, whose shape is consistent with that inferred from the GW binary BH (BBH) primaries, can be well described as a power-law with an index of $\alpha = 4.8^{+4.5}_{-2.8}$ plus a high-mass Gaussian component peaking at $\sim33^{+14}_{-9}\,M_\odot$. The NS mass spectrum could be shaped as a near flat distribution between $\sim1.0-2.1\,M_\odot$. The constrained NS maximum mass agrees with that inferred from NSs in our Galaxy. If GW190814 and GW200210 are NSBH mergers, the posterior results of the NS maximum mass would be always larger than $\sim2.5\,M_\odot$ and significantly deviate from that inferred in the Galactic NSs. The effective inspiral spin and effective precession spin of GW NSBH mergers are measured to potentially have near-zero distributions. The negligible spins for GW NSBH mergers imply that most events in the universe should be plunging events, which supports the standard isolated formation channel of NSBH binaries. More NSBH mergers to be discovered in the fourth observing run would help to more precisely model the population properties of cosmological NSBH mergers.

\end{abstract}


\section{Introduction} \label{sec:intro}

Neutron star mergers, including binary neutron star (BNS) and neutron star--black hole (NSBH) mergers, are the prime targeted gravitational-wave (GW) sources for the Advanced Laser Interferometer Gravitational-Wave Observatory \citep[LIGO][]{aasi2015}, Advanced Virgo \citep{acernese2015} and KAGRA \citep{aso2013} GW detectors. They have long been proposed to be progenitors of short-duration gamma-ray bursts \citep[sGRBs;][]{paczynski1986,paczynski1991,eichler1989,narayan1992,zhang2018} and kilonovae\footnote{sGRB jets from neutron star mergers in active galactic nucleus disks \citep[e.g.,][]{cheng1999,mckernan2020} would always be choked by the disk atmosphere \citep{perna2021,zhu2021neutrino1,zhu2021neutron}. Potential jet-cocoon and ejecta shock breakouts could produce fast-evolving optical transients and neutrino bursts \citep{zhu2021neutrino2,zhu2021neutrino1}. } \citep{li1998,metzger2010}. On 2017 August 17th, LIGO and Virgo detected the first GW signal from a BNS system \citep{abbott2017GW170817} which was confirmed to be associated with an sGRB \citep[GRB170817A;][]{abbott2017gravitational,goldstein2017,savchenko2017,zhangbb2018}, a fast-evolving ultraviolet-optical-infrared kilonova transient \citep[AT2017gfo;][]{abbott2017multimessenger,arcavi2017,coulter2017,drout2017,evans2017,kasliwal2017,pian2017,smartt2017,kilpatrick2017}, and a broadband off-axis jet afterglow \citep[e.g.,][]{margutti2017,troja2017,lazzati2018,lyman2018,ghirlanda2019}. The joint observations of the GW signal and associated electromagnetic (EM) counterparts from this BNS merger confirmed the long-hypothesized origin of sGRBs and kilonovae, and heralded the arrival of the GW-led multi-messenger era.

Compared with BNS mergers which would definitely eject a certain amount of materials to produce EM signals, some NSBH binaries may not tidally disrupt the NS component and, hence, would not make bright EM counterparts such as sGRBs and kilonovae\footnote{During the final merger phase for plunging NSBH binaries, some weak EM signals may be produced because of the charge and magnetic field carried by the NS \citep[e.g.,][]{zhang2019,dai2019,pan2019,dorazio2021,sridhar2021}.}. The tidal disruption probability of NSBH mergers and the brightness of NSBH EM signals are determined by BH mass, BH spin, NS mass, and NS equation of state \citep[EoS; e.g.,][]{belczynski2008,kyutoku2011,kyutoku2013,kyutoku2015,fernandez2015,kwagaguchi2015,kawaguchi2016,foucart2012,foucart2018,barbieri2019,kruger2020,fragione2021constraining,fragione2021BHNS,zhu2020,zhu2021no,zhu2021kilonova,tiwari2021,li2021,raaijmakers2021}. A NSBH merger tends to be a disrupted event and produces bright EM signals if it has a low mass BH with a high projected aligned-spin, and a low mass NS with a stiff EoS. The parameter space in which a NSBH merger can undergo tidal disruption may be very limited. Recently, LIGO-Virgo-KAGRA (LVK) Collaboration reported three high-confidence GWs from NSBH candidates, i.e., GW190814, GW200105\_162426 and GW200115\_042309 \citep{abbott2020GW190814,abbott2021observation,nitz2021}. In spite of many efforts for follow-up observations of these three events, no confirmed EM counterpart candidate has been identified \citep[e.g.,][]{anand2021,alexander2021,coughlin2020,gompertz2020,kasliwal2020,kilpatrick2021,page2020,thakur2020,dobie2021}. \cite{abbott2021observation,zhu2021no,fragione2021BHNS} showed that the parameter space of these GW candidates mostly lies outside the disrupted parameter region, so that these candidates are likely plunging events with a high probability. There have been many mysteries about NSBH binaries, such as the proportion of disrupted events in cosmological NSBH mergers, their cosmological contribution to the elements heavier than iron, the formation channel of NSBH binaries, and so on. A systemic research on the population properties of NSBH binaries can help us address these mysteries and unveil the nature of cosmological NSBH binaries.

LVK Collaboration has announced several GW candidates during the third observing run (O3) whose component masses were potentially consistent with originating from NSBH mergers \citep{abbott2020GW190814,abbott2021observation,abbott2021gwtc2,abbott2021gwtc21,abbott2021gwtc3}, although only some of these GW signals had a relatively low false alarm rate and a large astrophysical origin probability. In this work, by using a Bayesian framework to analyse the canonical results of these confirmed NSBH candidates reported by LVK Collaboration, we make a first step to investigate the population properties of GW NSBH mergers with the information of their mass and spin distributions.

\section{Method\label{sec:method}}

\subsection{Event Selection \label{sec:event}}

\begin{deluxetable*}{clrlrllrc}[htpb!]
\tablecaption{Source properties for potential NSBH candidates\label{tab:samples}}
\tablecolumns{5}
\tablewidth{0pt}
\tablehead{
\colhead{Event} &
\colhead{$\mathcal{M}/M_\odot$} &
\colhead{$m_1/M_\odot$} &
\colhead{$m_2/M_\odot$} &
\colhead{$\chi_{\rm eff}$} &
\colhead{$z$} &
\colhead{{FAR/${\rm yr}^{-1}$}} & 
\colhead{{$p_{\rm astro}$}}
}
\startdata
GW190426 & $2.41^{+0.08}_{-0.08}$ & $5.7^{+3.9}_{-2.3}$ & $1.5^{+0.8}_{-0.5}$ & $-0.03^{+0.32}_{-0.30}$ & $0.08^{+0.04}_{-0.03}$ & $9.1\times10^{-1}$ & 0.14\\
GW190917 & $3.7^{+0.2}_{-0.2}$ & $9.3^{+3.4}_{-4.4}$ & $2.1^{+1.5}_{-0.5}$ & $-0.11^{+0.24}_{-0.49}$ & $0.15^{+0.06}_{-0.06}$ & $6.6\times10^{-1}$ & 0.77 \\
GW191219 & $4.33^{+0.10}_{-0.15}$ & $31.6^{+1.8}_{-2.5}$ & $1.17^{+0.06}_{-0.05}$ & $0.00^{+0.07}_{-0.08}$ & $0.11^{+0.04}_{-0.03}$ & $4.0\times10^{0}$ & 0.82\\
GW200105 & $3.41^{+0.08}_{-0.07}$ & $8.9^{+1.1}_{-1.3}$ & $1.9^{+0.2}_{-0.2}$ & $-0.01^{+0.08}_{-0.12}$ & $0.06^{+0.02}_{-0.02}$ & $2.0\times10^{-1}$ & 0.36 \\
GW200115 & $2.42^{+0.05}_{-0.07}$ & $5.9^{+1.4}_{-2.1}$ & $1.4^{+0.6}_{-0.2}$ & $-0.14^{+0.17}_{-0.34}$ & $0.06^{+0.03}_{-0.02}$ & $3.0\times10^{-10}$ & $>0.99$ \\
\hline
GW190814 & $6.09^{+0.06}_{-0.06}$ & $23.2^{+1.1}_{-1.0}$ & $2.59^{+0.08}_{-0.09}$ & $-0.002^{+0.060}_{-0.061}$ & $0.05^{+0.009}_{-0.010}$ & $5.4\times10^{-12}$ & $>0.99$\\
GW200210 & $6.56^{+0.34}_{-0.38}$ & $24.5^{+8.9}_{-5.3}$ & $2.79^{+0.54}_{-0.48}$ & $0.03^{+0.25}_{-0.25}$ & $0.19^{+0.07}_{-0.06}$ & $1.2\times10^0$ & 0.54
\enddata
\tablecomments{The columns from left to right are [1] NSBH GW candidate; [2] chirp mass; [3] the mass of the primary component; [4] the mass of the secondary mass; [5] effective inspiral spin; [6] redshift; {[7] FAR in per yer; [8] probability of astrophysical origin.}}
\end{deluxetable*}

The LVK Collaboration has made public 7 NSBH candidates, including GW190426\_1522155, GW190814, GW190917\_114630, GW191219\_163120, GW200105\_162426, GW200115\_042309, and GW200210\_092254, which are respectively abbreviated as GW190426, GW190814, GW190917, GW191219, GW200105, GW200115, and GW200210, hereafter. We use the canonical posterior results of these NSBH candidates to investigate the population properties of GW NSBH mergers. 

{\cite{abbott2021populationO3b} used GWTC-3 GWs with a false alarm rate (FAR) $<0.25\,{\rm yr}^{-1}$ and an astrophysical probability $p_{\rm astro}>0.5$ to investigate the population properties of compact binary coalescences, whereas GW200105 with $p_{\rm astro}\sim0.36$ was also included in their study. For the binary BH (BBH) focused analyses, the events with ${\rm FAR}<1\,{\rm yr}^{-1}$ were also considered \citep{abbott2021populationO3b}. \cite{farr2015,roulet2020} combined the astrophysical probability from individual events to explore the population inference. In this work,} we simply assume that all of these NSBH candidates are {real signals and of} astrophysical origins. Among the 7 NSBH candidates, the secondary object of GW190814 and GW200210 could either be an NS or a BH, so it is uncertain whether they are NSBH mergers. Therefore, we collect the observations of GW candidates that were consistent with originating from NSBH binaries, including GW190426, GW190917, GW191219, GW200105, and GW200115, to derive the population properties in detail. These 5 NSBH candidates constitute  \textsc{group a}. We also separately take into account GW190814 and GW200210 as two other NSBH GW candidates to explore their influences on the population properties. We thus define \textsc{group b} to contain all 7 candidates. Because the secondary spins of these events are not well constrained by present GW observations, we focus on investigating the effective inspiral spin and the effective precessing spin. In order to study the distribution between the effective inspiral spin and the effective precessing spin for NSBH binary systems, {we adopt the published posterior samples inferred using the precession waveforms \texttt{IMRPhenomPv2} model \citep{hannam2014} for GW190426 and \texttt{IMRPhenomXPHM} model \citep{partten2021} for other GW events. The data releases were downloaded from the Gravitational Wave Transient Catalog (GWTC; \url{https://www.gw-openscience.org/eventapi/html/GWTC/})}. The posterior results of these NSBH candidates are summarized in Table \ref{tab:samples}.

\subsection{Population Models \label{sec:pop}} 

For Bayesian inference and modeling, by assuming that the BH mass ($m_1$) and NS mass ($m_2$) distributions are independent, we employ two typical BH mass distributions and three NS mass distributions. We consider to directly measure the distributions of the effective inspiral spin parameter ($\chi_{\rm eff}$) and the effective precession spin parameter ($\chi_{\rm p}$), in which the prior of the spin model is set as a bivariate Gaussian between $\chi_{\rm eff}$ and $\chi_{\rm p}$. In view that the O3 NSBH candidates are very close by, the redshift evolution can be neglected. Therefore, a \textsc{nonevolving} redshift model for NSBH mergers is implemented.

\subsubsection{Parameterized BH Mass Distribution}

Our simplest BH mass model is a \textsc{power-law} distribution with hard cutoffs at both minimum ($m_{1,{\rm min}}$) and maximum ($m_{1,{\rm max}}$) masses, i.e.,
\begin{equation}
\begin{split}
    \pi(m_1|\alpha,m_{\rm 1,min},m_{\rm 1,max}) \propto m_1^{-\alpha}, \\
    {\rm for}~m_{\rm 1,min}<m_1<m_{\rm 1,max}
\end{split}
\end{equation}
where $\alpha$ is the power-law index. This model, derived from \cite{ozel2010,fishbach2017,wysocki2019}, has been used to fit the BBH events \citep{abbott2019population,abbott2021populationO3a,abbott2021populationO3b}.

A group of BHs at $\sim 20 - 50\,M_\odot$ \citep{abbott2019population,abbott2021populationO3a,abbott2021populationO3b}, which cause a overdensity relative to a power law distribution, could be originated from pulsational pair-instability supernovae \citep{talbot2018}. Since the primary masses of GW190814, GW191219 and GW200210 are located in this range and much larger than those of other NSBH candidates, we also adopt a power-law distribution with a second Gaussian component in the high-mass region (\textsc{power-law + peak}) as our second BH mass distribution model. This model reads
\begin{equation}
\begin{split}
    \pi(m_1|\alpha,\mu_{m},\sigma_{m},m_{\rm 1,min},m_{\rm 1,max}) = \lambda_1\mathcal{N}(m_1|\mu_{m},\sigma_{m})/A_1 \\ + (1 - \lambda_1)m_{1}^{-\alpha}/A_2,~{\rm for}~m_{\rm 1,min}<m_1<m_{\rm 1,max}
\end{split}
\end{equation}
where $\mathcal{N}$ stands for Gaussian distribution, $\lambda_1$ is the fraction of primary BHs in the Gaussian component, $\mu_m$ is the mean of the Gaussian component, $\sigma_m$ is the standard deviation of the Gaussian component, and $A_1$ and $A_2$ are the normalization factors, respectively.

\subsubsection{Parameterized NS Mass Distribution}

The simplest NS mass model is defined as a \textsc{uniform} distribution between minimum ($m_{\rm 2,min}$) and maximum ($m_{\rm 2,max}$) masses, which has been used in \cite{landry2021,liyj2021}, i.e.,
\begin{equation}
\begin{split}
    \pi(m_2|m_{\rm 2,min},m_{\rm 2,max}) = 1/(m_{\rm 2,max} - m_{\rm 2,min}),\\{\rm for}~m_{\rm 2,min}<m_2<m_{\rm 2,max}.
\end{split}
\end{equation}

In view that the observationally derived mass distribution of Galactic BNS systems is approximately a normal distribution \citep{lattimer2012,kiziltan2013}, we also consider that the distribution of the NS mass in NSBH systems is a \textsc{single gaussian} distribution with a mean $\mu$ and a standard deviation $\sigma$. The model is expressed as
\begin{equation}
\begin{split}
    \pi(m_2|\mu,\sigma,m_{\rm 2,min},m_{\rm 2,max}) \propto \mathcal{N}(m_2|\mu,\sigma),\\{\rm for}~m_{\rm 2,min}<m_2<m_{\rm 2,max}.
\end{split}
\end{equation}

Because the masses of Galactic NSs can be well explained by a bimodal distribution \citep{antoniadis2016,alsing2018,farr2020,shao2020}, a \textsc{double gaussian} mass scenario is also considered as the prior of the NS mass distribution, which is taken to be
\begin{equation}
\begin{split}
    &\pi(m_2|\mu_1,\sigma_1,\mu_2,\sigma_2,\lambda_{2},m_{\rm 2,min},m_{\rm 2,max}) = \\ &\lambda_{2}\mathcal{N}(m_2|\mu_1,\sigma_1)/A_3 
    +(1 - \lambda_{2})\mathcal{N}(m_2|\mu_2,\sigma_2)/A_4,\\&~~~~~~~~~~~~~~~~~~~~~~~~~~~~{\rm for}~m_{\rm 2,min}<m_2<m_{\rm 2,max},
\end{split}
\end{equation}
where $\lambda_2$ is the fraction of NSs in the first low-mass Gaussian component, $\mu_{1}$ ($\mu_{2}$) and $\sigma_{1}$ ($\sigma_{2}$) are the mean and standard deviation of the first (second) Gaussian component, respectively, while $A_3$ and $A_3$ are the normalization factors.

\subsubsection{Parameterized Spin Distribution}

Motivated by \cite{miller2020,abbott2021populationO3a}, we parameterize the distributions of $\chi_{\rm eff}$ and $\chi_{\rm p}$ by assuming that their distributions are jointly described as a bivariate Gaussian, i.e., 
\begin{equation}
    \pi(\chi_{\rm eff},\chi_{\rm p}|\mu_{\rm eff},\sigma_{\rm eff},\mu_{\rm p},\sigma_{\rm p},\rho) \propto \mathcal{N}(\boldsymbol{\mu},\boldsymbol{\Sigma}),
\end{equation}
where $\boldsymbol{\mu} = (\mu_{\rm eff},\mu_{\rm p})$ is defined as the mean of $\chi_{\rm eff}$ and $\chi_{\rm p}$. 
\begin{equation}
    \boldsymbol{\Sigma} = 
    \begin{pmatrix}
    \sigma_{\rm eff}^2 & \rho\sigma_{\rm eff}\sigma_{\rm p} \\
    \rho\sigma_{\rm eff}\sigma_{\rm p} & \sigma_{\rm p}^2
    \end{pmatrix},
\end{equation}
is the covariance matrix of the spin distribution, where $\rho$ is the degree of correlation between $\chi_{\rm eff}$ and $\chi_{\rm p}$ while $\sigma_{\rm eff}$ and $\sigma_{\rm p}$ are assumed to be standard deviations of the $\chi_{\rm eff}$ and $\chi_{\rm p}$ distributions.

\subsection{Hierarchical Population Model}

We perform a hierarchical Bayesian approach, marginalizing over the properties of individual events and the number of expected NSBH detections during O3, to measure the given population parameters for the distributions of BH mass ($\boldsymbol{\Lambda}_{m_1}$), NS mass ($\boldsymbol{\Lambda}_{m_2}$), and spin ($\boldsymbol{\Lambda_\chi}$). Given a set of data $\boldsymbol{d}_{i}$ from $N_{\rm det}$ NSBH GW detections, the likelihood as a function of given combined population hyperparameters $\boldsymbol{\Lambda}$ can be expressed as \citep[e.g.,][]{fishbach2018,mandel2019,vitale2020,abbott2021populationO3a,abbott2021populationO3b,farah2021}
\begin{equation}
\label{equ:likelihood}
    \mathcal{L}(\{\boldsymbol{d}\}|\boldsymbol{\Lambda}) \propto  \prod^{N_{\rm det}}_{i = 1} \frac{\int\mathcal{L}(\boldsymbol{d}_i|\boldsymbol\theta)\pi(\boldsymbol{\theta}|\boldsymbol{\Lambda})d\boldsymbol{\theta}}{\xi(\boldsymbol{\Lambda})},
\end{equation}
where $\boldsymbol{\theta} = (m_1,m_2,\chi_{\rm eff},\chi_{\rm p})$ is the event parameters, $\mathcal{L}(\boldsymbol{d}_i|\boldsymbol{\theta})$ is the single-event likelihood, $\pi(\boldsymbol{\theta}|\boldsymbol{\Lambda}) = \pi(m_1|\boldsymbol{\Lambda}_{m_1})\pi(m_2|\boldsymbol{\Lambda}_{m_2})\pi(\chi_{\rm eff},\chi_{\rm p}|\boldsymbol{\Lambda_{\chi}})$ is the combined prior, and $\xi(\boldsymbol{\Lambda})$ is the detection fraction. Using the posterior
samples of NSBH mergers described in Section \ref{sec:event} to evaluate $\mathcal{L}(\boldsymbol{d}_i|\boldsymbol{\theta})$, Equation (\ref{equ:likelihood}) can be further given by \citep[][]{abbott2021populationO3a,abbott2021populationO3b}
\begin{equation}
    \mathcal{L}(\{\boldsymbol{d}\}|\boldsymbol{\Lambda})\propto\prod^{N_{\rm det}}_{i = 1}\frac{1}{\xi(\boldsymbol{\Lambda})}\left< \frac{\pi(\boldsymbol{\theta}|\boldsymbol{\Lambda})}{\pi_{\varnothing}(\boldsymbol{\theta})} \right>_{\rm samples},
\end{equation}
where $\pi_{\varnothing}(\boldsymbol{\theta})$ is the default prior adopted for initial parameter estimation.

The detection fraction is
\begin{equation}
\begin{split}
    \xi(\boldsymbol{\Lambda}) &= \int P_{\rm det}(\boldsymbol{\theta})\pi(\boldsymbol{\theta}|\boldsymbol{\Lambda})d\boldsymbol{\theta}, \\
    &\approx \int P_{\rm det}(m_1,m_2)\pi(m_1|\boldsymbol{\Lambda}_{m_1})\pi(m_2|\boldsymbol{\Lambda}_{m_2})dm_1dm_2,
\end{split}
\end{equation}
where $P_{\rm det}$ is the probability that a NSBH event can be detected. $\chi_{\rm eff}$ and $\chi_{\rm p}$ have less influence on the detection probability of a NSBH event \citep[e.g.,][]{zhu2021kilonova} so that we ignore the effect of them on the detection probability. We then simulate $P_{\rm det}(m_1,m_2)$ based on the method introduced in \cite{abbott2021populationO3a}. 

In our simulation, we employ \texttt{dynesty} sampler \citep{speagle2020} to evaluate the likelihoods for population models while \texttt{GWPopulation} package \citep{talbot2019} is used for the implementation of the likelihoods. Table \ref{tab:Priors} describes the priors adopted for each of our hyperparameters $\boldsymbol{\Lambda}$.

\begin{deluxetable*}{ccccccc}
\tablecaption{Priors of hierarchical Bayesian inference\label{tab:Priors}}
\tablewidth{0pt}
\tablehead{
\multirow{2}{*}{Parameters} &  \multicolumn6c{BH mass distribution priors} \\ \cline{2-7} & \multicolumn3c{\textsc{power-law}}  & \multicolumn3c{\textsc{power-law+peak}}
}
\startdata
$m_{\rm 1,min}/M_\odot$ & \multicolumn3c{U$(3,8)$} & \multicolumn3c{U$(3,8)$} \\
$m_{\rm 1,max}/M_\odot$ & \multicolumn3c{U$(20,60)$} & \multicolumn3c{U$(20,60)$} \\
$\alpha$ & \multicolumn3c{U$(-4,12)$} & \multicolumn3c{U$(-4,12)$} \\
$\lambda_1$ & \multicolumn3c{$-$} & \multicolumn3c{U$(0,1)$} \\
$\mu_{m}/M_\odot$ & \multicolumn3c{$-$} & \multicolumn3c{U$(20,50)$} \\
$\sigma_{m}/M_\odot$ & \multicolumn3c{$-$}  & \multicolumn3c{U$(0.01,10)$} \\
Constraint & \multicolumn3c{$-$} & \multicolumn3c{$m_{\rm 1,min}<\mu_{m}<m_{\rm 1,max}$} \\ \hline
 & \multicolumn{6}{c}{NS mass distribution priors} \\ \cline{2-7} 
 & \multicolumn2c{\textsc{uniform}} & \multicolumn2c{\textsc{single gaussian}} & \multicolumn2c{\textsc{double gaussian}} \\ \hline
$m_{\rm 2,min}/M_\odot$ & \multicolumn2c{U$(1,1.5)$} & \multicolumn2c{U$(1,1.5)$} & \multicolumn2c{U$(1,1.5)$} \\
$m_{\rm 2,max}/M_\odot$ & \multicolumn2c{U$(1.5,3.5)$} & \multicolumn2c{U$(1.5,3.5)$} & \multicolumn2c{U$(1.5,3.5)$} \\
$\mu/M_\odot$ & \multicolumn2c{$-$} & \multicolumn2c{U$(1,3.5)$} & \multicolumn2c{$-$} \\
$\sigma/M_\odot$ & \multicolumn2c{$-$} & \multicolumn2c{U$(0.01,1.3)$} & \multicolumn2c{$-$} \\
$\lambda_2$ & \multicolumn2c{$-$} & \multicolumn2c{$-$} & \multicolumn2c{U$(0,0.1)$} \\
$\mu_1/M_\odot$ & \multicolumn2c{$-$} & \multicolumn2c{$-$} & \multicolumn2c{U$(1,3.5)$} \\
$\sigma_1/M_\odot$ & \multicolumn2c{$-$} & \multicolumn2c{$-$} & \multicolumn2c{U$(0.01,1.3)$} \\
$\mu_2/M_\odot$ & \multicolumn2c{$-$} & \multicolumn2c{$-$} & \multicolumn2c{U$(1,3.5)$} \\
$\sigma_2/M_\odot$ & \multicolumn2c{$-$} & \multicolumn2c{$-$} & \multicolumn2c{U$(0.01,1.3)$} \\
Constraint & \multicolumn2c{$-$} & \multicolumn2c{$m_{\rm 2,min}<\mu<m_{\rm 2,max}$} & \multicolumn2c{$m_{\rm 2,min}<\mu_{1}<\mu_{2}<m_{\rm 2,max}$} \\ \hline
 & \multicolumn{6}{c}{\textsc{gaussian} spin distribution priors} \\ \hline
$\mu_{\rm eff}$ & \multicolumn{6}{c}{U$(-1,1)$} \\
$\sigma_{\rm eff}$ & \multicolumn{6}{c}{U$(0.01,1)$} \\
$\mu_{\rm p}$ & \multicolumn{6}{c}{U$(0.01,1)$} \\
$\sigma_{\rm p}$ & \multicolumn{6}{c}{U$(0.01,1)$} \\
$\rho$ & \multicolumn{6}{c}{U$(-0.75,0.75)$}
\enddata
\tablecomments{U represents uniform distribution.}
\end{deluxetable*}

\section{Results}

\begin{deluxetable*}{ccccc}
\tablecaption{The Bayes factors of BH and NS mass distribution models\label{tab:Bayes}}
\tablewidth{0pt}
\tablehead{
\multirow{2}{*}{Events} & \multirow{2}{*}{BH mass distribution} & \multicolumn3c{NS mass distribution} \\ \cline{3-5} & & \colhead{\textsc{uniform}} & \colhead{\textsc{single gaussian}} & \colhead{\textsc{double gaussian}}
}
\startdata
\textsc{group a} & \textsc{power-law} & $1.0\pm0.4\,(0.0\pm0.2)$ & $0.4\pm0.1\,(-0.4\pm0.1)$ & $0.2\pm0.1\,(-0.8\pm0.2)$ \\ \cline{2-5} 
(excluding GW190814 \& GW200210) & \textsc{power-law+peak} & $2.5\pm0.3\,(0.4\pm0.1)$ & $1.2\pm0.2\,(0.1\pm0.1)$ & $0.7\pm0.1\,(-0.2
\pm0.1)$ \\ \hline
\textsc{group b} & \textsc{power-law} & $1.0\pm0.2\,(0.0\pm0.1)$ & $0.7\pm0.2\,(-0.1\pm0.1)$ & $1.4\pm0.5\,(0.1\pm0.1)$ \\\cline{2-5} 
(including GW190814 \& GW200210) & \textsc{power-law+peak} & $2.5\pm0.6\,(0.4\pm0.1)$ & $1.5\pm0.5\,(0.2\pm0.1)$ & $4.5\pm1.3\,(0.7\pm0.1)$ \\
\enddata
\tablecomments{The values of Bayes factor $\mathcal{B}$ {(log Bayes factor $\log_{10}\mathcal{B}$ in brackets)} for each NSBH group are relative to the evidence of the \textsc{power-law} BH mass distribution and the \textsc{uniform} NS mass distribution. }
\end{deluxetable*}

As shown in Section \ref{sec:method}, there are totally twelve synthesis prior models, i.e., two groups of NSBH candidates (\textsc{group a} and \textsc{group b}) $\times$ two BH mass distribution models (\textsc{power-law} and \textsc{power-law+peak}) $\times$ three NS mass distribution models (\textsc{uniform}, \textsc{single gaussian} and \textsc{double gaussian}) $\times$ one redshift evolution model (\textsc{nonevolving}). We provide Bayes factors $\mathcal{B}$ {(log Bayes factors $\log_{10}\mathcal{B}$)} comparing different models in Table \ref{tab:Bayes}.

\subsection{BH Mass Distribution}

\begin{figure}[tbp] 
    \centering
	\includegraphics[width = 1\linewidth , trim = 85 310 90 20, clip]{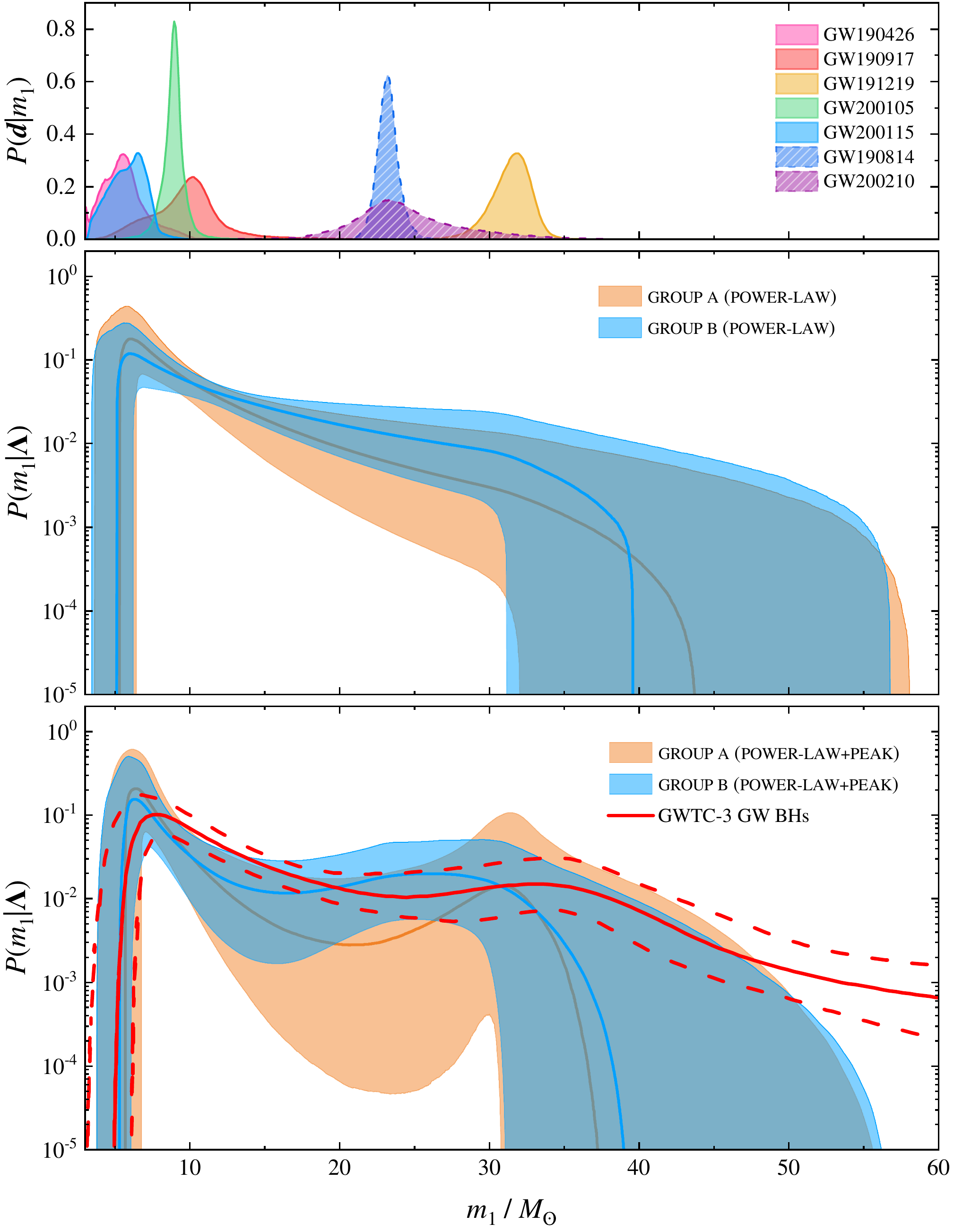}
    \caption{Top panel: one-dimensional posterior distributions for the masses of the BHs in O3 GW NSBH candidates. Middle and bottom panels: the BH mass distributions inferred for the \textsc{power-law} and \textsc{power-law+peak} models. Orange (blue) solid line and shaded region represent the median and $90\%$ confidence interval obtained by adopting \textsc{group a} (\textsc{group b}), respectively. Red lines are the GWTC-3 BBH primary mass distribution \citep{abbott2021populationO3b} with the median (solid) and $90\%$ confidence interval (dashed). }\label{fig:BHmass}
\end{figure}

Figure \ref{fig:BHmass} displays the posterior distribution of the primary BH masses for O3 NSBH candidates. Taken \textsc{uniform} model as a fixed model for the NS mass distribution, the BH mass distribution inferred by the \textsc{power-law} and \textsc{power-law+peak} models for both groups of NSBH events are also plotted in Figure \ref{fig:BHmass} as examples. 

When we fit the \textsc{power-law} BH mass distribution model to the data of \textsc{group a}, the power-law index is $\alpha = 2.7^{+2.0}_{-1.5}$ between the sharp low-mass cutoff $m_{\rm 1,min} = 5.3^{+1.1}_{-1.7}\,M_\odot$ and high-mass cutoff $m_{\rm 1,max} = 44^{+14}_{-12}\,M_\odot$. Since the primary masses of GW190814 and GW200210 are larger than those of other NSBHs expect GW191219, if we include them in the data, the mass model would have a relatively shallower slope with $\alpha = 1.7^{+1.3}_{-1.3}$ from $m_{\rm 1,min} = 5.1^{+1.1}_{-1.7}\,M_\odot$ to $m_{\rm 1,max} = 40^{+17}_{-8}\,M_\odot$. $m_{\rm 1,max}$ depends on the priors, but the lower bound on $m_{\rm 1,max}$ is driven by the precise mass measurement for $\sim32\,M_\odot$ primary of GW191219.

The \textsc{power-law} model may be disfavored to explain the mass distribution of BHs in NSBH systems due to the lack of NSBH GW detections with a primary BH mass in the range of $\sim 12-20\,M_\odot$, so that a more complicated BH mass distribution model with the consideration of bimodal components is needed. For different NS mass distribution models and groups of NSBH events, the results of \sout{log} Bayes factor (see Table \ref{tab:Bayes}) reveal that the \textsc{power-law+peak} model of the BH mass distribution has a moderate preference over the \textsc{power-law} model by {a Bayes factor of} $\sim2:1-4:1$ ($\log_{10}\mathcal{B}\sim0.4-0.6$). For \textsc{group a} (\textsc{group b}), we find a power-law slope of $\alpha = 4.8^{+4.5}_{-2.8}$ ($\alpha = 4.0^{+5.1}_{-2.7}$), supplemented by a Gaussian peak at $\mu_m = 33^{+14}_{-9}\,M_\odot$ ($\mu_m = 27^{+15}_{-6}\,M_\odot$), between minimum mass $m_{\rm 1,min} = 5.7^{+1.1}_{-1.3}\,M_\odot$ ($m_{\rm 1,min} = 5.6^{+1.1}_{-1.5}\,M_\odot$) and maximum mass $m_{\rm 1,max} =38^{+19}_{-7}\,M_\odot$ ($m_{\rm 1,max} = 39^{+18}_{-9}\,M_\odot$). Comparing with those inferred from the \textsc{power-law} model, the BH mass distribution derived by the \textsc{power-law+peak} model would have steeper slopes, but obtain consistent minimum and maximum masses.

\cite{abbott2021populationO3b} showed that the \textsc{power-law+peak} model can also well fit to the primary mass of GWTC-3 BBH GW events. They found that the primary mass distribution of BBH, plotted in the bottom panel of Figure \ref{fig:BHmass}, would have a power-law slope of $\alpha = 3.4^{+0.58}_{-0.49}$ with a Gaussian peak at $\mu_m = 34^{+2.3}_{-3.8}\,M_\odot$. The comparison of the primary mass distributions by the GW detections indicate that the primary components of cosmological NSBH and BBH mergers could have similar minimum mass distributions and similar power-law slopes. The maximum primary mass of NSBH mergers is much lower than that of BBH mergers because of the present limited number of detections for high-mass NSBH mergers. Constrained by \textsc{group a}, the primary BH of GW191219 dominates the high-mass component, which would result in a similar mean of the Gaussian component and a similar probability distribution compared with those of BBH mergers. The similar shapes of the primary BH mass distribution between GW NSBH and BBH mergers indicate that the NSBH mergers reported in GWTC-3 are likely credible and plausibly have an astrophysical origin. When we include GW190814 and GW200210, the mean of the Gaussian component is less massive than that of GWTC-3 BBH mergers.

\subsection{NS Mass Distribution}

\begin{figure}[tbp] 
    \centering
	\includegraphics[width = 1\linewidth , trim = 85 30 90 20, clip]{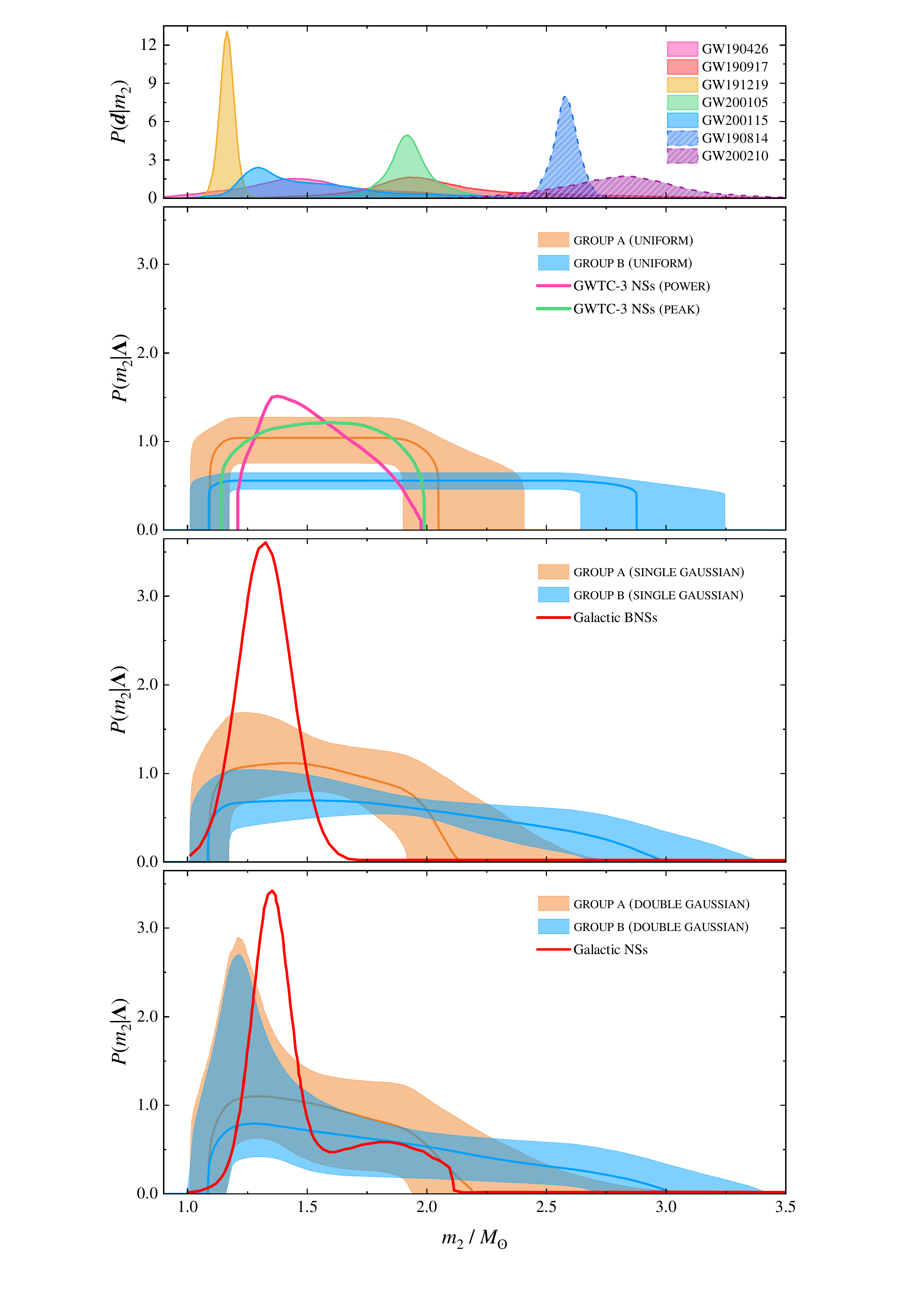}
    \caption{Top panel: one-dimensional posterior distributions for the masses of the NSs in O3 GW NSBH candidates. Following three panels: the NS mass distributions inferred for the \textsc{uniform}, \textsc{single gaussian} and \textsc{double gaussian} models. The orange (blue) solid lines and shaded regions represent the median with the $90\%$ confidence interval by adopting the data of \textsc{group a} (\textsc{group b}), respectively. Derived by \cite{abbott2021populationO3b}, the pink and green lines are the median mass distribution of the NSs in GW BNS and NSBH mergers using \textsc{power} and \textsc{peak} models. The red lines shown in the \textsc{single gaussian} and \textsc{double gaussian} models are the NS masses of galactic BNSs \citep{kiziltan2013} and galactic NSs \citep{farr2020}.}\label{fig:NSmass}
\end{figure}

The top panel of Figure \ref{fig:NSmass} shows the posterior distribution of the secondary NS masses for O3 NSBH candidates. Furthermore, by setting \textsc{power-law+peak} model as a fiducial model for BH mass distribution, we plot the medians and 90\% confidence intervals of the inferred NS mass distributions for the \textsc{uniform}, \textsc{single gaussian}, and \textsc{double gaussian} models in Figure \ref{fig:NSmass}.

Given the data of \textsc{group a}, three models exhibit a similar fitting result for the NS mass distribution. All models show a consistent minimum NS mass, $m_{\rm 2,min} = 1.088^{+0.086}_{-0.076}\,M_\odot$. The maximum NS mass inferred by the \textsc{uniform} model is $m_{\rm 2,max} = 2.04^{+0.35}_{-0.15}\,M_\odot$, while $m_{\rm 2,max}$ would be a little higher and broader, i.e., $m_{\rm 2,max} = 2.1^{+1.0}_{-0.2}\,M_\odot$ and $m_{\rm 2,max} =2.2^{+1.1}_{-0.3}\,M_\odot$, if one respectively considers the \textsc{single gaussian} and \textsc{double gaussian} NS mass models. Regardless of which prior models one adopts, the final NS mass distributions look like a uniform distribution between minimum and maximum masses. 

Using \textsc{group b} as the observational input, {by a Bayes factor of $\sim2:1-3:1$ $(\log_{10}\mathcal{B}\sim0.3-0.5)$}, the \textsc{double gaussian} model provides a better fit than the \textsc{uniform} and \textsc{single gaussian} models to the shape of the NS mass distribution. The fitting results of \textsc{double gaussian} model are $\mu_1 = 1.30^{+0.57}_{-0.21}\,M_\odot$, $\sigma_1 = 0.61^{+0.76}_{-0.52}\,M_\odot$, $\mu_2 = 2.04^{+0.85}_{-0.72}\,M_\odot$, and $\sigma_2 = 0.94^{+0.50}_{-0.71}\,M_\odot$. However, the inferred NS mass distribution does not show apparent bimodal structures but presents a linear decline to the maximum mass after the peak. Due to the presence of two other events (i.e., GW190814 and GW200210) with a high-mass secondary, the maximum mass would be much higher with a larger uncertainty compared with the result inferred by the input of \textsc{group a}, i.e., $m_{\rm 2,max} = 3.02^{+0.40}_{-0.33}\,M_\odot$.

\begin{figure}
    \centering
	\includegraphics[width = 1\linewidth , trim = 75 30 80 60, clip]{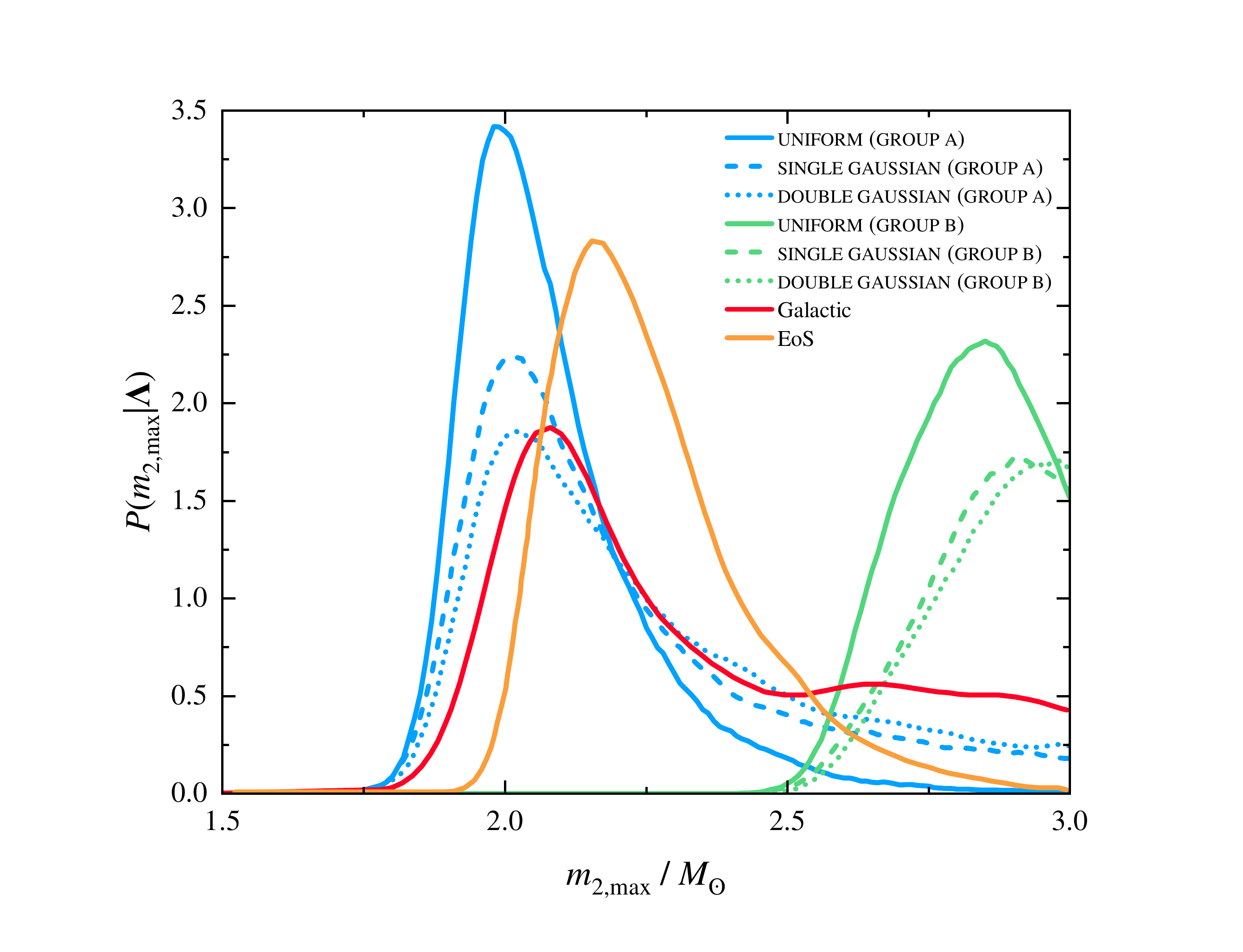}
    \caption{Maximum mass distribution for the NSs in GW NSBH binaries. Blue (green) solid, dashed and dotted lines show the inferred maximum mass distribution by adopting \textsc{uniform}, \textsc{single gaussian}, and \textsc{double gaussian} models for the data of \textsc{group a} (\textsc{group b}), respectively. The red and orange lines represent the maximum mass derived from the Galactic NS population in \cite{farr2020} and the maximum TOV mass from the EoS predicted by \cite{landry2020}. }\label{fig:maximumNSmass}
\end{figure}

In view of the lack of observations for Galactic NSBH binaries, we briefly compare our results of NS masses with the mass distributions of the galactic BNSs \citep{kiziltan2013}, Galactic NSs \citep{farr2020}, and GW NSs reported in GWTC-3 \citep{abbott2021populationO3b}. As shown in Figure \ref{fig:NSmass}, in comparison with the Galactic BNSs and NSs that have a narrow mass distribution in the low-mass region, the mass distribution of NSs observed in GW NSBH mergers is broader and has greater prevalence for high-mass NSs. This may be because NSBH systems with high-mass NSs could merge early, and hence, predominantly low-mass NSs remain observed in our Galaxy. Furthermore, GW NSs reported by \cite{abbott2021populationO3b} and our fitting NSs in GW NSBH binaries similarly show a broad, relatively flat mass distribution. 

The Galactic NS population \citep{farr2020} and the maximum Tolman-Oppenheimer-Volkov (TOV) mass predicted by  \cite{landry2020} gave the value of $m_{\rm 2,max} = 2.3^{+0.8}_{-0.3}\,M_\odot$ and $m_{\rm TOV} = 2.2^{+0.4}_{-0.2}\,M_\odot$, respectively. These constrained masses have a good consistency with our maximum masses of NSs inferred by the data of \textsc{group a}. As illustrated in Figure \ref{fig:maximumNSmass}, if we also take GW190814 and GW200219 into consideration, the simulated maximum mass would be always larger than $\sim2.5\,M_\odot$ and peak at $\sim2.7-3.0\,M_\odot$. The overlap between the maximum NS mass in GW NSBH binaries and the maximum masses observed in the Galaxy would be limited. The secondary maximum mass in the GW NSBH mergers would significantly deviate from that inferred from the NSs in our Galaxy.

\subsection{Spin Distribution}

\begin{figure*}
    \centering
	\includegraphics[width = 0.49\linewidth , trim = 80 30 80 65, clip]{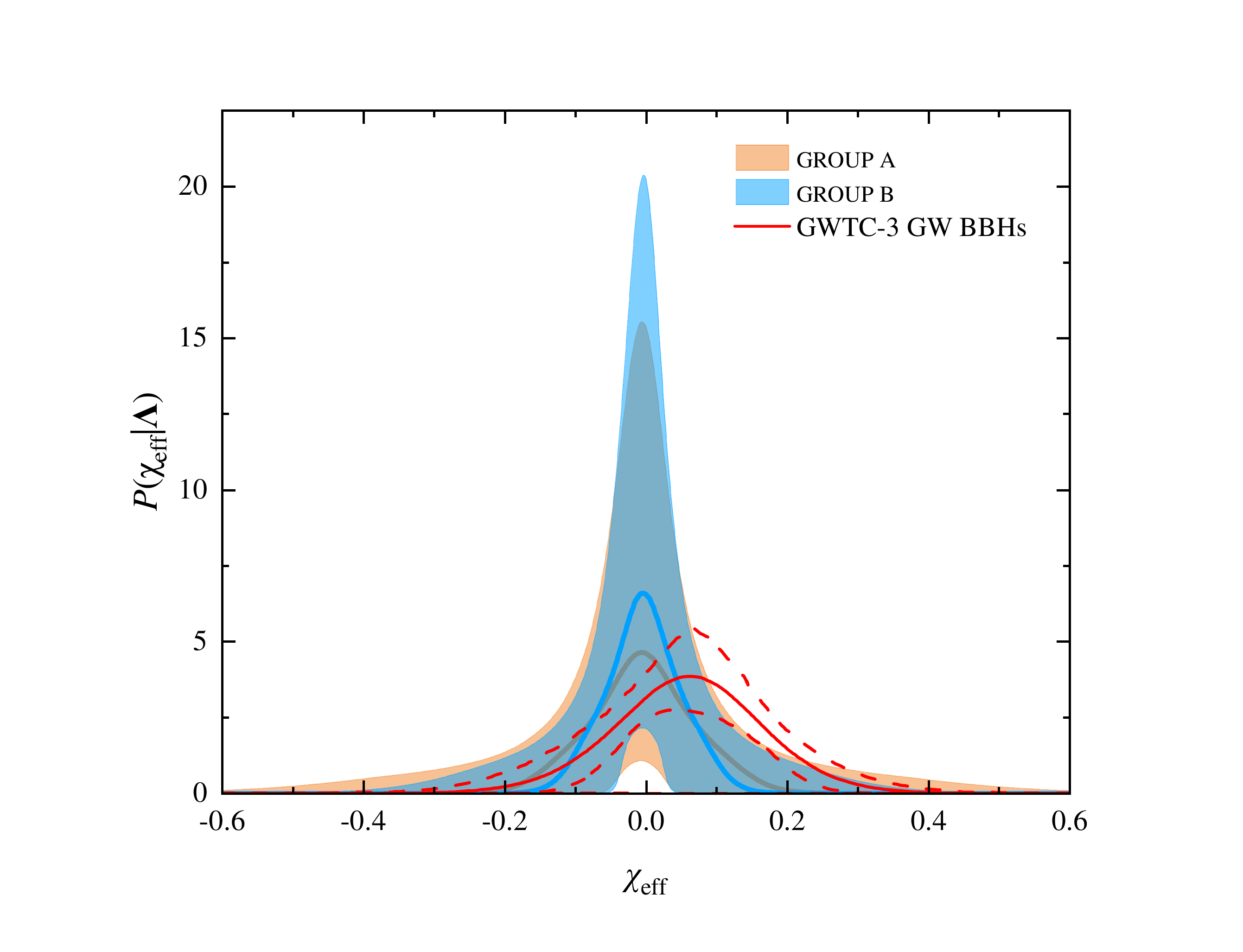}
	\includegraphics[width = 0.49\linewidth , trim = 80 30 80 65, clip]{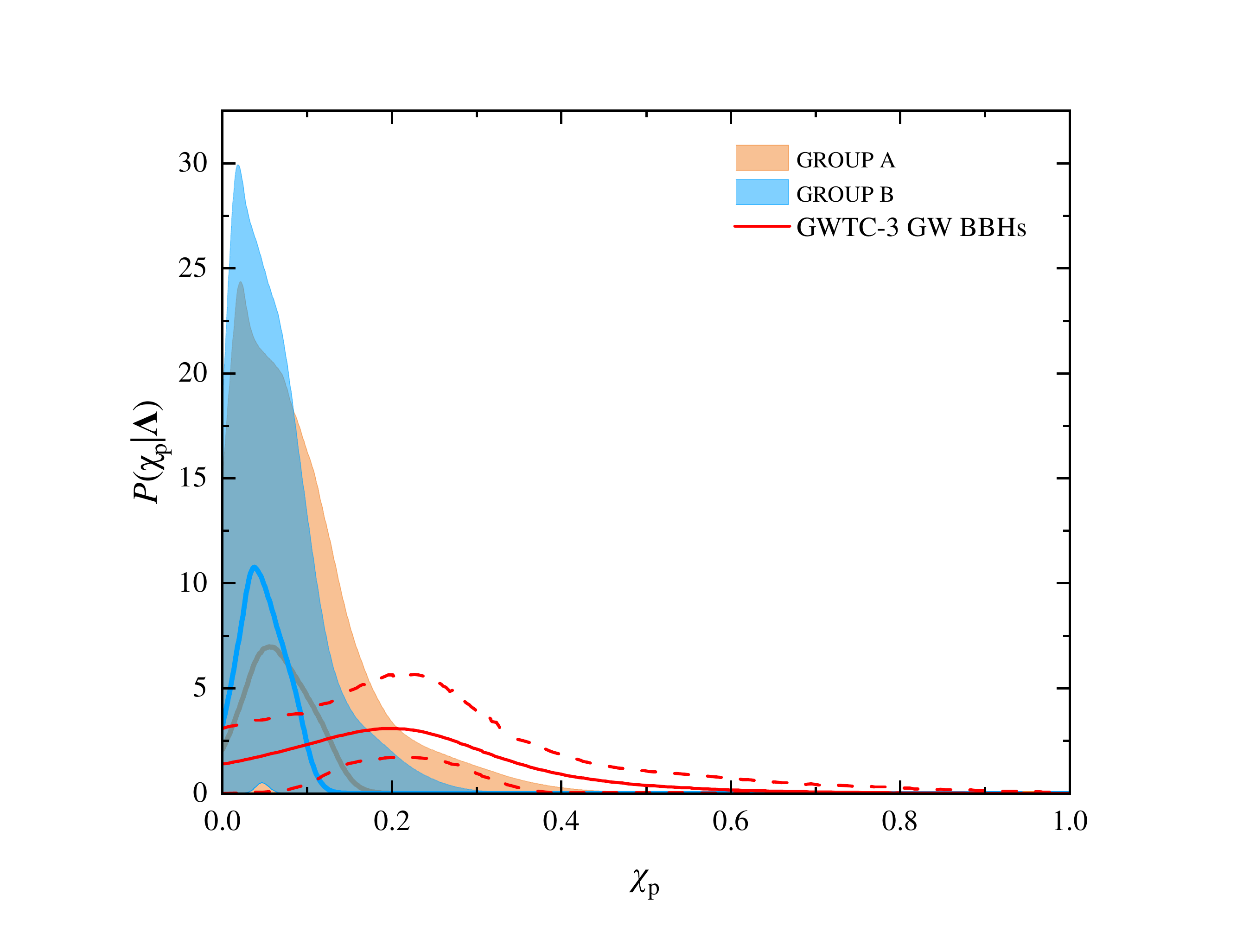}
    \caption{Population distributions for $\chi_{\rm eff}$ (left panel) and $\chi_{\rm p}$ (right panel) of NSBH systems. Orange and blue shaded regions show the central $90\%$ credible bounds using \textsc{group a} and \textsc{group b}, while the solid lines show the median posterior prediction. The red solid and dashed lines mark the central $50\%$ and $90\%$ posterior credible regions for O3 GW BBH systems from \cite{abbott2021populationO3b}, respectively}\label{fig:spin}
\end{figure*}

Figure \ref{fig:spin} illustrates our example constraints on the $\chi_{\rm eff}$ and $\chi_{\rm p}$ distributions under the models of the \textsc{power-law+peak} BH mass distribution, the \textsc{uniform} NS mass distribution, and the \textsc{gaussian} spin distribution. Regardless of the prior models that we choose, our constrained results reveal that both $\chi_{\rm eff}$ and $\chi_{\rm p}$ of NSBH mergers could have near-zero distributions. However, due to the limited number of detections in O3, the spin distributions display large uncertainties.

For the data of \textsc{group a}, the posterior distributions of the median and the standard deviation of $\chi_{\rm eff}$ ($\chi_{\rm p}$) are $\mu_{\rm eff} = -0.004^{+0.046}_{-0.053}$ ($\mu_{\rm p} = 0.064^{+0.084}_{-0.049}$) and $\sigma_{\rm eff} = 0.07^{+0.18}_{-0.06}$ ($\sigma_{\rm p} = 0.03^{+0.10}_{-0.02}$), respectively. \textsc{group b} data further support a negligible spin distribution. In this case, we obtain $\mu_{\rm eff} = -0.003^{+0.030}_{-0.035}$ ($\mu_{\rm p} = 0.046^{+0.056}_{-0.033}$) and $\sigma_{\rm eff} = 0.05^{+0.11}_{-0.04}$ ($\sigma_{\rm p} = 0.024^{+0.061}_{-0.013}$). {All of these individual GW candidates, except for GW200115, have near-zero $\chi_{\rm eff}$ and $\chi_{\rm p}$. It is expected that removing some of the marginal events would not significantly affect the spin population distributions.} Because the spin of the BH component contributes to most of $\chi_{\rm eff}$ and $\chi_{\rm p}$, the measurements of negligible spin distribution indicate that BHs in cosmological NSBH systems could have a low-spin population distribution.

In Figure \ref{fig:spin}, we also show the comparison with the resulting spin distributions of BBH mergers made with GWTC-3 \citep{abbott2021populationO3b} using the \textsc{gaussian} spin model. The spin measurements for BBH mergers suggested an effective inspiral spin distribution of non-vanishing width centered at $\chi_{\rm eff} = 0.06^{+0.04}_{-0.04}$ and a narrow precession spin distribution centered around $\chi_{\rm p} \approx 0.2$. By contrast, the population distributions of $\chi_{\rm eff}$ and $\chi_{\rm p}$ for NSBH systems are lower than those of BBH systems.

\subsection{Event Rate}

GWTC-3 reported a NSBH merger rate of $7.4 - 320.0\,{\rm Gpc}^{-3}\,{\rm yr}^{-1}$. In our work, the inferred merger rate mainly depends on the adopted input of observational data. By setting \textsc{powerlaw+peak} as the BH mass model and \textsc{uniform} as the NS mass model, for \textsc{group a}, different mass models give a consistent NSBH rate merger, which is $13.2 - 64.9\,{\rm Gpc}^{-3}\,{\rm yr}^{-1}$. If GW190814 and GW200210 were NSBH mergers, the uncertainty of the NSBH rate merger would be lower, i.e., $13.8 - 53.0\,{\rm yr}^{-1}$. {The inferred event rate does not increase since the low-mass end of the BH mass spectrum is relatively unaffected by these two additional events for the \textsc{powerlaw+peak} model as they would be picked up by the high-mass peak component.}

\section{Discussion}

\subsection{Tidal Disruption Probability}

\begin{figure}[htpb]
    \centering
    \includegraphics[width = 1\linewidth , trim = 50 20 70 20, clip]{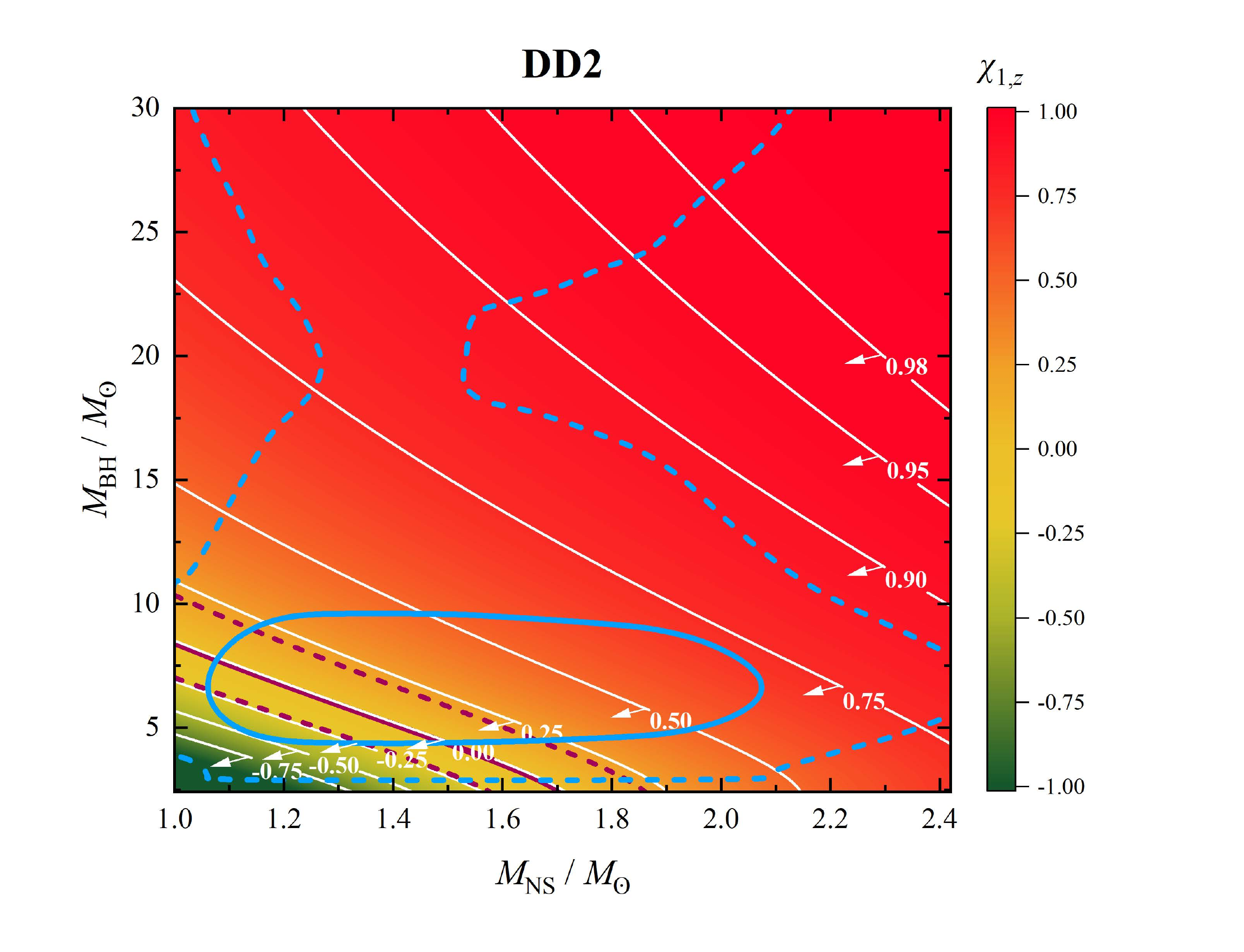}
    \caption{The source-frame mass parameter space where tidal disruption can occur by considering the specific NS EoS DD2. We mark several values of primary BH spin along the orbital angular momentum from $\chi_{1,z} = - 0.75$ to $\chi_{1,z} = 0.98$ as solid white lines. For a specific $\chi_{1,z}$, the NSBH mergers with component masses located at the bottom left parameter space (denoted by the direction of the arrows) can allow tidal disruptions to occur. The blue solid and dashed lines represent $50\%$ and $90\%$ source-frame masses distributions of inferred population results for O3 GW NSBH mergers. The dark purple solid and dashed lines represent median and $90\%$ distributions of $\chi_{1,z}$, respectively.}
    \label{fig:tidal}
\end{figure}

We calculate the amount of total baryon mass after NSBH mergers, which is mainly determined by BH mass, BH aligned spin, NS mass, and NS EoS, to judge whether or not tidal disruption happens using an empirical model presented by \cite{foucart2018}. We generate cosmological NSBH merger events based on the posterior results obtained by the \textsc{power-law+peak} BH mass distribution and the \textsc{single gaussian} NS mass distribution for the observational input of \textsc{group a}. For each posterior population sample, we simulate 1000 events including the system parameters of BH mass, NS mass, and effective spin. It is plausibly expected that most NSs would have near-zero spins before NSBH mergers owing to the spindown process via magnetic dipole radiation \citep[e.g.,][]{manchester2005,oslowski2011}. The primary BH spin along the orbital angular momentum can be thus estimated as $\chi_{1,z} \approx (m_1 + m_2)\chi_{\rm eff}/m_1$. A NS EoS of DD2 \citep{typel2010} is adopted, since this EoS is one of the stiffest EoSs constrained by GW170817 \citep{abbott2018GW170817,abbott2019properties,dietrich2020}. 

Figure \ref{fig:tidal} shows the parameter space where the NS can be tidally disrupted. The 50\% and 90\% distributions of BH mass, NS mass, and BH aligned spin for our simulated NSBH mergers are also plotted in Figure \ref{fig:tidal}. Because the simulated BHs have a common aligned spin in the range of $-0.25\lesssim\chi_{1,z}\lesssim0.25$, the mass space that allows NS tidal disruption and produce bright EM signals would require $m_1\lesssim7\,M_\odot$ and $m_2\lesssim1.5\,M_\odot$. However, most of our simulated NSBH mergers inferred from the GW observations have BHs and NSs located outside of the tidal disruption mass space. This indicates that plunging events would account for a large fractional of cosmological NSBH mergers.

\subsection{Implications for the Formation Channel}

Among O3 NSBH candidates, \cite{abbott2021observation} reported that the BH component of GW200115 could have a misaligned spin and an orbital precession. Recently, many works in the literature, e.g., \cite{broekgaarden2021formation,fragione2021impact,gompertz2021,zhu2021}, presented that a moderate or strong natal kick for the BH or the NS is required in order to produce the observed misalignment angle of GW200115. On the other hand, applying alternative astrophysically motivated priors to GW200115, \cite{mandel2021} constrained the BH spin to be centered at zero. It would thus result in a more negligible population spin distribution.

The most promising formation scenario for NSBH binaries is isolated binary evolution \citep[e.g.,][]{broekgaarden2021,shao2021}. Furthermore, a small fraction of NSBH binaries are believed to result from dynamical evolution \citep[e.g.,][]{clausen2013,ye2020}. In the standard scenario for merging NSBH formation through isolated binary evolution, the primary (initially more massive star) evolves off the main sequence, initiates mass transfer onto the secondary, and finally collapses to form a BH before the common-envelope phase. During this process, the primary evolves in a wide orbit in which the tides are too weak to spin it up. Additionally, the angular momentum content of the primary is reduced by stripping off its outer layers due to stellar winds and mass transfer via the first Lagrangian point onto its companion. Furthermore, under the assumption of efficient angular momentum transport within the star, predicted by the Tayler-Spruit dynamo \citep{spruit2002} or its revised version by \cite{fuller2019}, the spin of the first-born BH is found to be small \citep{qin2018,belz2020,drozda2020}. Our present results for constraints on the spin population properties of GW NSBH mergers would strongly support the standard scenario for the formation of cosmological NSBH binaries.

Alternatively, the first-born compact object in NSBH mergers could be a NS. \cite{romangaza2021} recently claimed that the fraction of the systems with a first-born NS with different supernova engines is $\sim 10\%$. This scenario would lead to a group of NSBH mergers with a unique spin distribution \citep{hu2022}, which has not been discovered by present GW detections. \cite{zhu2021kilonova} further found that the brightness of kilonova is strongly dependent on the spin magnitude of the BH in the NSBH mergers. The possible energy injection from BH-torus would also be affected by the primary BH spin \citep[e.g.,][]{ma2018,qi2021}. Therefore, we suggest a detailed investigation of the BH spin-dependent parameter space in the near future.

\section{Conclusions}

In this work, we analyse the canonical results of these confirmed NSBH candidates reported by LVK Collaboration by employing a Bayesian framework to study the population properties of GW NHBH mergers. A power-law with an Gaussian peak model can well explain the NSBH primary mass distribution. The posterior distribution of the power-law index ($\alpha = 4.8^{+4.5}_{-2.8}$), the minimum mass ($m_{\rm 1,min} = 5.3^{+1.1}_{-1.7}$), and the mean of the Gaussian feature ($\mu_{m} = 33^{+14}_{-9}\,M_\odot$) have similar properties with those of GWTC-3 BBH primary mass distribution. The mass distribution of the NS component is consistent with a uniform distribution between $\sim1.0-2.1\,M_\odot$, similar to the constraint on the masses of NSs in GW binaries \citep{landry2021,abbott2021populationO3b}. The maximum mass distribution derived by GW NSBH mergers agrees with that inferred from the NSs in our Galaxy. If these NSBH candidates reported by LVK Collaboration are of an astrophysical origin, the event rate of NSBH mergers would be $13.2 - 64.9
\,{\rm Gpc}^{-3}\,{\rm yr}^{-1}$. 

If GW190814 and GW200210 are NSBH mergers, the BH mass spectrum can be also fitted by a power-law distribution with a high-mass Gaussian component. A \textsc{double gaussian} model is supported to account for the NS mass distribution. However, the inferred NS mass distribution does not show apparent bimodal structures, which performs a linear decline to the maximum mass after the peak of $\sim1.3\,M_\odot$. The secondary maximum mass in the GW NSBH mergers would significantly increase and result in an apparent deviation from that inferred from the NSs in our Galaxy. The event rate of NSBH mergers would change to $13.8 - 53.0\,{\rm Gpc}^{-3}\,{\rm yr}^{-1}$. 

Different from GWTC-3 BBH systems that show non-vanishing distributions of spins, GW NSBH systems display near-zero distributions for both effective inspiral spin and effective precession spin. Because the NS component makes an insignificant contribution to the spin of the system, the negligible spin distributions for NSBH populations plausibly indicate that most BHs in the cosmological NSBH systems would have a low spin. This result would support the standard isolated formation channel of NSBH binaries. Since the primary BHs likely have low spins, plunging events would be the dominant population for NSBH mergers and hence no bright EM counterparts are expected for most of NSBH mergers. 

{We have assumed that GWTC-3 NSBH candidates are real signals of astrophysical origins when we analyse the population properties of GW NSBH mergers. Interestingly, (1) the similar shapes of the primary BH mass distribution between GW NSBH and BBH mergers, (2) the consistent shapes of the NS mass distribution between NS mergers and NSBH mergers, (3) near-zero spin distributions for NSBH populations which are supported by the standard isolated formation channel of NSBH binaries, indicate that the NSBH mergers reported in GWTC-3 are likely credible and plausibly real signals with astrophysical origins. In the fourth observation run, more discovered NSBH mergers would help to more precisely model the population properties of cosmological NSBH mergers.}

\acknowledgments
We thank {an anonymous referee for valuable suggestions and} Yong Shao for helpful comments. We acknowledge the Atlas cluster computing team at AEI Hannover. J.P.Z is partially supported by the National Science Foundation of China under Grant No.~11721303 and the National Basic Research Program of China under grant No.~2014CB845800. Y.Q. is supported by the Doctoral research start-up funding of Anhui Normal University and by the National Natural Science Foundation of China under Grant No.~12192221. H.G. is supported by the National Natural Science Foundation of China under Grant No.~11690024, 12021003, 11633001. Z.J.C. is supported by the National Natural Science Foundation of China under Grant No.~11920101003, 12021003 and CAS Project for Young Scientists in Basic Research YSBR-006.

\software{\texttt{Python}, \url{https://www.python.org}; \texttt{Matlab}, \url{https://www.mathworks.com}; \texttt{GWPopulation} \citep{talbot2019}; \texttt{LALSuite} \citep{lalsuite}  }

\bibliography{NSBH}{}
\bibliographystyle{aasjournal}

\end{document}